\documentclass{article}
\usepackage{spconf,amsmath,graphicx}
\usepackage{amssymb,mathtools}
\usepackage{multirow}
\usepackage{relsize}
\usepackage{xcolor}
\usepackage{booktabs}
\usepackage{cite}
\usepackage{hyperref}

\usepackage{amsmath,amsfonts,bm}

\def\eqref#1{equation~\ref{#1}}
\def\1{\bm{1}}

\def\vx{{\bm{x}}}
\def\vy{{\bm{y}}}
\def\vz{{\bm{z}}}

\DeclareMathAlphabet{\mathsfit}{\encodingdefault}{\sfdefault}{m}{sl}
\SetMathAlphabet{\mathsfit}{bold}{\encodingdefault}{\sfdefault}{bx}{n}

\def\gL{{\mathcal{L}}}

\title{Efficient Knowledge Distillation for RNN-Transducer Models}

\name{Sankaran Panchapagesan, Daniel S. Park,
 Chung-Cheng Chiu, Yuan Shangguan\sthanks{Work performed
	while at Google}, Qiao Liang,}
\nameplus{Alexander Gruenstein}
\address{Google LLC, U.S.A., ${}^*$Facebook Inc, U.S.A.\\
	\texttt{\small\{panchi, danielspark, chungchengc, wildstone, alexgru\}@google.com, ${}^*$yuansg@fb.com}
 	}

\begin{document}
%\ninept
%
\maketitle
\begin{abstract}
Knowledge Distillation is an effective method of transferring knowledge from a large model to a smaller model. Distillation can be viewed as a type of model compression, and has played an important role for on-device ASR applications. In this paper, we develop a distillation method for RNN-Transducer (RNN-T) models, a popular end-to-end neural network architecture for streaming speech recognition. Our proposed distillation loss is simple and efficient, and uses only the ``y" and ``blank" posterior probabilities from the RNN-T output probability lattice. We study the effectiveness of the proposed approach in improving the accuracy of sparse RNN-T models obtained by gradually pruning a larger uncompressed model, which also serves as the teacher during distillation. With distillation of 60\% and 90\% sparse multi-domain RNN-T models, we obtain WER reductions of 4.3\% and 12.1\% respectively, on a noisy FarField eval set. We also present results of experiments on LibriSpeech, where the introduction of the distillation loss yields a 4.8\% relative WER reduction on the test-other dataset for a small Conformer model.
\end{abstract}
\section{Introduction}
\label{sec:intro}

Knowledge Distillation \cite{bucilua2006model, Hinton15} is known to help improve ASR model accuracy \cite{PangSainath2018, Mosner2019}.
Distillation is a type of model compression, which is important for on-device applications.
However, there is no established approach for distilling RNN transducer (RNN-T) models \cite{Graves12}, a popular end-to-end neural network architecture for streaming ASR \cite{he2019streaming}.
The goal of this work is to develop an effective approach for RNN-T model distillation.

In Knowledge Distillation a {\em teacher} model is used to help guide the training of a \textit{student} model \cite{Hinton15}. The teacher could be a single larger model, or an ensemble of models \cite{chebotar2016distilling}, while the student is typically a smaller model that fits within the compute constraints to run in real time during inference.
A distillation loss between teacher and student predictions: 
$\displaystyle \gL_\text{distill} = d(\bm{\widetilde{P}}, \bm{P})$,
where tilde denotes teacher, is used to encourage the student to imitate the teacher. Hence, knowledge distillation is a type of model compression, where a student model with fewer parameters attempts to capture the behavior of the teacher model(s) with more parameters.

Typically, for classifier models, the KL divergence between teacher and student posteriors is used:
\begin{equation}
\displaystyle \gL_\text{distill} = \text{KL}(\bm{\widetilde{P}}, \bm{P}) = \sum_{k} \bm{\widetilde{P}}(k|\vx) \ln \frac{\bm{\widetilde{P}}(k|\vx)}{\bm{{P}}(k|\vx)} \,.
\end{equation}
Since the teacher model is usually fixed, the cross entropy between teacher and student predictions can be used. It is also common to add a Temperature parameter, which is tuned \cite{Hinton15}.
The overall training loss on labeled data is typically an interpolation between the base loss and the distillation loss. 
The distillation loss above can be straightforwardly generalized to a teacher-student pair of hybrid \cite{hinton2012deep}, CTC \cite{graves2006connectionist} or LAS \cite{chan2016listen} models. But it is more subtle to apply it to RNN-T models, where the loss is given by the negative log conditional probability obtained by summing over all possible alignments of the encoder-produced features and the output tokens.

In this work, we introduce a distillation loss for RNN-T networks in equation (\ref{eq:Lsimple}), given by the sum of KL divergences between teacher and student conditional probabilities over the entire RNN-T output probability lattice. As explained further in section \ref{sec:rnnt-distillation}, for efficiency, the loss is computed as the sum of KL divergences between coarse-grained conditional probabilities, that lump together the probabilities for any tokens  other than the distinguished ``y" token (the correct label at the particular output step), and the ``blank" token (which shifts the alignment in the time direction).

For sparse RNN-T models trained on a multi-domain dataset similar to that described in \cite{narayanan2018toward}, we are able to improve performance by introducing the distillation loss during sparsification,  leading to relative word-error-rate (WER) reductions of 4.3\% and 12.1\% for 60\% and 90\% sparse models, respectively, on a noisy FarField test set. We also experiment with Conformer models \cite{gulati2020conformer} on LibriSpeech, and find a 4.8\% relative WER reduction on the test-other dataset performance by a small Conformer model upon introducing a distillation loss against a larger Conformer model.

\subsection{Related Work}

Distillation \cite{bucilua2006model, Hinton15} has been applied to speech recognition in varying contexts. Ideas similar to distillation
%where soft labels obtained from a teacher model are used to train student networks,
have a long history of being applied to speech tasks, e.g., in \cite{Pearce2000TheAE} for spoken digit recognition and in \cite{Yu_etal_2013} for speaker adaptation. Distillation has been applied to compression of ASR models including hybrid \cite{chebotar2016distilling, watanabe2017student, lu2017knowledge, fukuda2017efficient, Mosner2019}, CTC \cite{sak2015acoustic, takashima2018investigation, takashima2019investigation, kurata2018improved, kurata2019guiding} and LAS models \cite{PangSainath2018}.

We propose a distillation scheme for RNN-T models, where unlike the models that previous distillation efforts have been focused on, the loss is computed over all possible alignments of the encoder features and output tokens. We then use the proposed distillation loss to better compress RNN-T models.

\vspace{-0.05in}
\section{RNN-T Model}
\label{sec:rnnt-model}
\vspace{-0.05in}

In this section, we briefly summarize the loss for the RNN Transducer \cite{Graves12} and set up some notation. The RNN-T training loss is given by the negative log posterior probability:
\begin{equation}
\displaystyle \gL = - \ln \bm{{P}}(\vy^*|\vx) \,,
\end{equation}
for target token sequence $\vy^*$ and input features $\vx$.

\begin{figure}[t]
\centering
\includegraphics[width=4.5cm]{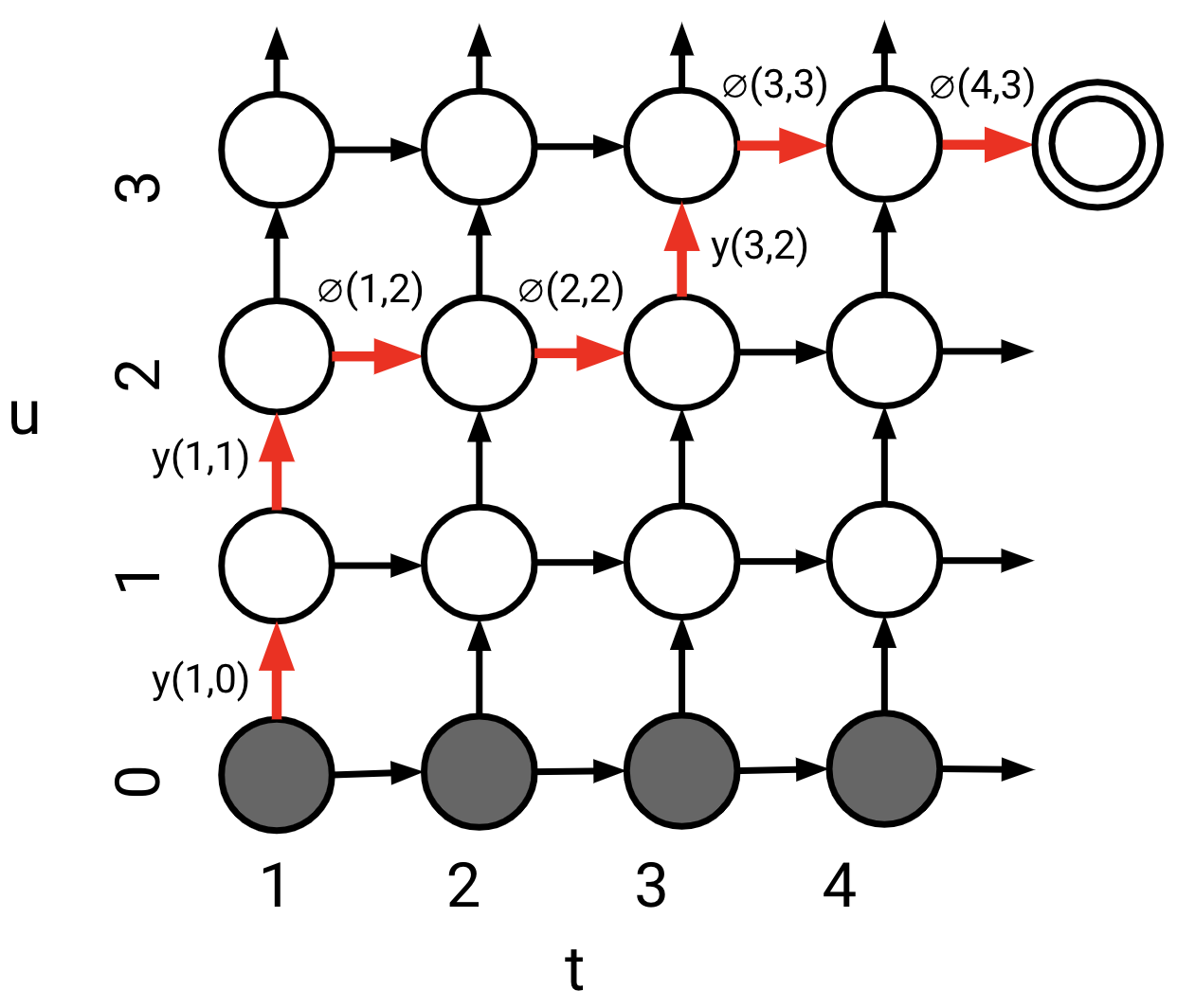}
\vspace{-0.1in}
\caption{RNN-T output probability lattice, following \cite{Graves12}.}
\label{fig:probability-lattice}
\vspace{-0.1in}
\end{figure}

The probability $\bm{{P}}(\vy^*|\vx)$ is computed via a $T \times U$ output probability lattice, depicted in figure \ref{fig:probability-lattice} (following \cite{Graves12}), where $T$ is the length of the encoder output feature sequence, denoted $\vz = \vz_{[1:T]}$, and $U$ is the length of the target output token sequence (including the end-of-sentence token), denoted $\vy = \vy_{[0:(U-1)]}$.
At each lattice node $(t, u)$, where $t \in \{1, 2, \dots, T\}$ and $u \in \{0, 1, \dots, (U-1)\}$, features from the encoder and prediction networks are used by the joint network and softmax layer to compute a probability distribution $\bm{P}(k|t,u)$ over output tokens $k$, including a special ``blank" token $\varnothing$.
Each path through the lattice from the origin to the final node represents an alignment between $\vz$ and $\vy$. During a horizontal transition from $(t, u)$, the model consumes a feature vector $\vz_{t+1}$ and emits the $\varnothing$ token with probability $\varnothing(t,u) \equiv \bm{P}(\varnothing|t,u)$, while during a vertical transition from $(t, u)$, the model emits the next output token $y_{u+1}$ with probability $y(t,u) \equiv \bm{P}(y_{u+1}|t,u)$ without consuming any more features. The probability for each alignment path through the lattice is given by the product of transition probabilities along the path. $\bm{{P}}(\vy^*|\vx)$ is obtained by summing the probabilities over all alignment paths.
In \cite{Graves12}, this sum was shown to be computed efficiently using a forward-backward algorithm utilizing the ``y" and ``blank" probabilities.

\vspace{-0.05in}
\section{RNN-T Distillation}
\label{sec:rnnt-distillation}
\vspace{-0.05in}

We have reviewed that the conditional probability used for the RNN-T loss is computed by summing over all possible encoder-output alignments, represented by paths on the output probability lattice. The canonical choice for a distillation loss, then, would be the sum of KL divergences between the token probability distributions of the teacher and student models over the lattice:
\begin{align}
\displaystyle \gL &= \sum_{u,t} \sum_k \bm{\widetilde{P}}(k|t,u) \ln \frac{\bm{\widetilde{P}}(k|t,u)}{\bm{{P}}(k|t,u)} \,. \label{eq:grid_loss_1}
\end{align}
The minimum of this loss is achieved when the entire output probability lattices of the teacher and student networks agree.

Assuming that the probability distributions are computed from feature vectors $h$ defined at each lattice point:
\begin{equation}
\bm{{P}}(k|h,t) = {\exp(h^{k}_{t,u}) \Big/ \sum_{k} \exp(h^k_{t,u})}\,,
\end{equation}
the derivative of the loss (\ref{eq:grid_loss_1}) lends itself to the expression:
\begin{equation}
{\partial \mathcal{L} / \partial h^k_{t,u}} = - \left(\bm{\widetilde{P}}(k|t,u) - \bm{{P}}(k|t,u) \right) \,.
\end{equation}

Denoting the output symbol vocabulary size $K$, the sum in the loss is of size $O(U \times T \times K)$. Moreover, the loss requires a memory usage of $2 \times U \times T \times K$ per utterance, which ranges from being extremely expensive to unpractical, depending on $K$. For example, with $U = 100$, $T = 500$ and $K = 4000$, the memory requirement per utterance is $1.6$ GB, assuming the losses are of float type. For a batch size of $1024$ utterances, the per-batch memory usage becomes $1.6$ TB.

Instead of distilling the output distribution over all tokens, a more efficient distillation approach is to focus on learning the transition behavior of the teacher. As the transition behavior on the RNN-T lattice is dictated by the ``y" and ``blank" posterior matrices, we propose to make the distillation loss (\ref{eq:grid_loss_1}) less expensive by focusing on these posteriors. We coarse grain the $K$-dimensional logits down to three dimensions by only keeping track of the conditional probabilities for ``y," ``blank" and the remainder labels, i.e.,
\begin{align}
\bm{P}_y(t,u) &= \bm{P}(y_{u+1}|t,u) = y(t,u)\,, \nonumber \\
\bm{P}_\varnothing(t,u) &= \bm{P}(\varnothing|t,u) = \varnothing(t,u)\,, \\
\bm{P}_r(t,u) &= 1 - \bm{P}_y(t,u) - \bm{P}_\varnothing(t,u) \nonumber \,,
\end{align}
and similarly for $\bm{\widetilde{P}}$. We can then define the distillation loss to be the KL divergence between these conditional probabilities:
\begin{equation}
  \gL_\text{distill} = \sum_{u,t} \sum_{\ell \in \{y, \varnothing, r \}} \bm{\widetilde{P}}_\ell(t,u) \ln \frac{\bm{\widetilde{P}}_\ell(t,u)}{\bm{{P}}_\ell(t,u)} \,.
  \label{eq:Lsimple}
\end{equation}
The computation and memory cost of this loss is $O(U \times T)$ compared to $O(U \times T \times K)$ for equation (\ref{eq:grid_loss_1}).

For distilled models, the overall training loss is a linear combination of the base RNN-T loss and the distillation loss:
\begin{equation}
\gL = \gL_\text{rnnt} + \beta \gL_\text{distill}
\label{eq:overall_loss}
\end{equation}
where $\beta$, the distillation loss weight, is tuned.

\vspace{-0.1in}
\section{Experiments}
\label{sec:experiments}
\vspace{-0.07in}

We performed two sets of experiments to assess the effectiveness of the proposed RNN-T distillation loss. The first set of experiments were conducted on Librispeech data with Conformer-RNN-T models \cite{gulati2020conformer} where the RNN-T distillation loss is used in a conventional teacher-student training setting. The second set of experiments were carried out on a large multi-domain dataset with sparse LSTM RNN-T models \cite{Shangguan2020}, where distillation is used alongside sparsification.

\vspace{-0.05in}
\subsection{Librispeech Experiments}
\label{sec:librispeec-expts}
\vspace{-0.05in}

We examined the effect of introducing a distillation loss on the LibriSpeech \cite{librispeech} dataset. We used 80-dimensional filter bank coefficients of the utterances as single-channel input features. We experimented with a modified version of the 118.8M-parameter ``Conformer L" model and the 10.3M-parameter ``Conformer S" model introduced in \cite{gulati2020conformer} as the ASR networks for LibriSpeech, where the batch-normalization layers are replaced with group-norm layers \cite{wu2018group} with the group number set to 2. The training parameters for the Conformer L and S models were set to be the same as those used in the original work \cite{gulati2020conformer}, trained with batch-size 2048 and a transformer learning-rate schedule \cite{vaswani2017attention} with 10k warm-up steps. We did not use language models.

\begin{table}[h!]
  \vskip -0.05in
  \caption{WERs(\%) from LibriSpeech experiments.}
  \vskip 0.05in
  \label{t:librispeech}
  \centering
  \small
  \resizebox{\columnwidth}{!}{%
  \begin{tabular}{lcccc}
    \toprule
    \bfseries Method & \bfseries dev & \bfseries dev-other & \bfseries test & \bfseries test-other \\
    \midrule
    \bfseries Baselines \\
    \quad  Conformer L
    & 1.9 & 4.4 & 2.0 & 4.5 \\
    \quad Conformer S
    & 2.4 & 6.2 & 2.7 & 6.3 \\
    \midrule
    \bfseries Teacher-Student \\
    \quad  Conformer S + Distillation Loss
    & 2.4 & 6.1 & 2.7 & 6.0 \\
    \bottomrule
  \end{tabular}
  }
\end{table}

We trained a Conformer S model on LibriSpeech with and without a distillation loss from a teacher Conformer L model. We tuned the loss parameter $\beta$ over three different values (1e-2, 1e-3 and 1e-4), while keeping all other training parameters unchanged and found that best performance is achieved at $\beta = \text{1e-3}$. An adaptive SpecAugment \cite{specaugment, specaugment2} policy with two frequency masks with mask parameter $F = 27$, and ten time masks with maximum time-mask ratio $p_S = 0.05$ has been used to augment the input, which was shared by the teacher and student model. The performance of the trained network is recorded in table \ref{t:librispeech}. Training with a distillation loss resulted in performance improvement on the noisy dev and test sets, with a 4.8\% relative WER improvement on the test-other set.

\vspace{-0.05in}
\subsection{Multi-domain Experiments}
\label{sec:sparse-expts}
\vspace{-0.05in}

We applied distillation to the training of sparse multi-domain RNN-T models, typically aimed at on-device applications.
Models were trained on a large multi-domain dataset similar to that described in \cite{narayanan2018toward}, where the domains include Search and FarField. The shared input of the teacher and student models are augmented using SpecAugment \cite{specaugment, specaugment2} and multi-style training \cite{kim2017generation}. The architecture of our uncompressed 0\% sparse RNN-T model, also the teacher model for distillation, is similar to that described in \cite{Shangguan2020}, and is as follows.  The RNN-T encoder and decoder respectively contain 8 and 2 LSTM layers with Coupled Input and Forget Gates (CIFG), with each CIFG LSTM layer composed of  2048 hidden units and a projection size of 640. The joint network is a feedforward network with 640 units, followed by a softmax layer with output size 4096, which is the number of modeled word pieces. The uncompressed model has 96M parameters.

We trained compressed sparse models of two different sizes. For a given sparse model, all LSTM cell input and recurrent matrices are pruned to one sparsity level, while LSTM projection matrices are pruned to a different sparsity level.
For the two sparse models we trained, these two sparsity levels are set to 60\%-60\% and 90\%-70\%, respectively. The two models have 41M and 14M parameters, respectively. We therefore broadly refer to the two models as being {\it 60\% sparse} and {\it 90\% sparse}, reflecting their sparsity levels rounded to the nearest multiple of 10. The joint network size of the 90\% sparse model is reduced to 320 from 640. Both models are initialized as much as possible from the uncompressed model, all layers for the 60\% sparse model, and all layers except the joint network and output layer for the 90\% sparse model.

 The pruning method used was similar to that described in \cite{ZhuGupta2018, PangSainath2018, Shangguan2020}, with a gradual, cubic polynomial sparsity schedule. A saliency criterion based on the product of the weight and the corresponding gradient of the loss function, was used to update the pruning mask that identifies weights to prune out \cite{YangGradientPruning2019, lee2019snip}. Block sparsity with block size $16 \times 1$ was used for more efficient computation on modern processors \cite{narang2017blocksparse}. The 60\% sparse and 90\% sparse models were trained separately with different pruning schedules. The pruning schedule for the 60\%/90\% sparse model starts at 50k/100k steps and ends at 100k/200k steps, respectively. 
Since the sparse models were obtained by gradual sparsification of the uncompressed model, distillation can be viewed as a form of regularization.

We selected a Search set and a FarField set as Dev sets, and a noisy FarField set as the Test set.
Models were trained for around 650k steps, and were evaluated on the dev sets at checkpoints spaced 10k steps apart starting from 500k steps. For each model, one checkpoint with best performance on the dev sets was selected and evaluated on the test set. 

The distillation loss weight, $\beta$ in equation (\ref{eq:overall_loss}) was tuned by training models with different weights in \{1e-4, 3e-4, 1e-3, 3e-3, 1e-2\}, and selecting the best weight according to the dev set WERs.
We found that optimal performance, summarized in table \ref{table:new-results}, is achieved by setting $\beta = 1\text{e-3}$ for both models.
As seen in table \ref{table:new-results}, the proposed approach yielded WER reductions of 4.3\% and 12.1\% respectively, for 60\% and 90\% sparse models, on the noisy FarField eval set.

\vspace{-0.1in}
% Formatted to save some lines
\begin{table}[h!]
  \caption{WER(\%) and WERR(\%) results with distillation of 60\% and 90\% sparse RNN-T Models with distillation loss weight $\beta$ = 1e-3. WERR shown in parentheses.}
  \label{table:new-results}
  \vskip 0.1in
  \centering
  \resizebox{\columnwidth}{!}{%
  \begin{tabular}{lcccccc}
%   \begin{tabular}{lrlrlrl}
    \toprule
    \bfseries Model/Method & 
    \multicolumn{4}{c}{\bfseries Dev} &
    \multicolumn{2}{c}{\bfseries Test} \\
    \cmidrule(r){2-5} \cmidrule(r){6-7}
    \bfseries (\# Params) &
    \multicolumn{2}{c}{\bfseries Search} &
    \multicolumn{2}{c}{\bfseries FarField} &
    \multicolumn{2}{c}{\bfseries FarField} \\
    \midrule
    \bfseries Uncompressed (96M)
    & ~6.0 & --~ & ~4.6 & --~ & ~6.6 & --~ \\
    \midrule
    \bfseries 60\% Sparse (41M)
    & ~6.6 & --~ & ~5.0 & --~ & ~7.0 & --~ \\
    ~+ Distillation Loss (WERR)
    & \bfseries ~6.3 &(4.5)~& ~5.0 &(0.0)~& \bfseries ~6.7 & (4.3)~\\
    \midrule
    \bfseries 90\% Sparse (14M)
    & ~9.2 & --~ & ~7.4 & --~ & ~9.1 & --~ \\
    ~+ Distillation Loss (WERR)
    & \bfseries ~8.7 & (5.4)~ &
    \bfseries ~6.7 & (9.5)~ &
    \bfseries ~8.0 &  (12.1)~ \\
    \bottomrule
  \end{tabular}
  }
\end{table}
\vspace{-0.2in}

\section{Discussion and future work}
\label{sec:discussion}
\vspace{-0.05in}

The distillation loss introduced in this work has recently been incorporated into other applications. We briefly summarize these results and discuss future directions of research.
\smallskip

\noindent\textbf{Universal ASR:} 
In \cite{yu2020universal}, jointly training a full-context (non-streaming) ASR model and a streaming ASR model, with the two models sharing parameters, is shown to significantly improve both accuracy and latency of the streaming model, thereby obtaining state-of-the-art streaming results on Librispeech and MultiDomain data. An important component of the training process in that work was in-place distillation from the full-context model to the streaming model, using the distillation loss proposed in this paper. In an ablation study with a ContextNet \cite{Han2020ContextNet} model, they found that removing in-place distillation resulted in 20\% worse WER on Librispeech TestOther and increased the median latency from 40ms to 120ms (see \cite{yu2020universal} for more details).
\smallskip

\noindent\textbf{Dynamic Sparsity ASR:}
In \cite{Wu2021DynamicSparsity}, dynamic sparsity neural networks (DSNN) are proposed, which, once trained, can be operated at different sparsity levels.
There, the introduction of our distillation loss for in-place distillation from non-sparse to sparse models was found to yield WER reductions of 0.1-0.4 absolute on the Search test set for models with different levels of sparsity, with a WERR of 5\% for a 70\% sparse model.
\smallskip

\noindent\textbf{Future Directions:}
Encouraging consistency between teacher and student outputs for different augmentations of the input features to the teacher and student - e.g. clean features for the teacher and noisy features for the student, as in \cite{Mosner2019}, could improve model robustness and give additional accuracy improvements.
While we have used only ``y" and ``blank" logits for the distillation loss in equation (\ref{eq:Lsimple}), our coarse-grained KL loss could also be applied to use the top-K logits, similar to \cite{Mosner2019}.
A \textit{Temperature} parameter could be included in the loss function and tuned.
Encoder output distillation using L2 or another loss, could help provide additional guidance from the teacher to the student and yield improvements.

\vspace{-0.05in}
\section{Summary}
\label{sec:summary}
\vspace{-0.05in}

In this paper, we have proposed a novel and efficient distillation loss for RNN-T models. The loss comes from a "lattice-based" approach that aims to bring the entire teacher and student output probability lattices closer to each other, and uses only the "y" and "blank" posteriors in a simplified KL divergence loss.
The loss has been successfully applied to distillation of sparse RNN-T models trained on multi-domain data.  The proposed  approach yielded WER reductions of 4.3\% and 12.1\% respectively, for 60\% and 90\% sparse models, on a noisy FarField eval set.
For a small Conformer model trained on LibriSpeech data, the introduction of the distillation loss yielded a 4.8\% relative WER reduction on the test-other dataset.
The proposed distillation loss has been incorporated successfully in other recent work, yielding WER improvements in dynamic-sparsity neural networks \cite{Wu2021DynamicSparsity}, and yielding significant improvements in both WER and latency for the streaming-mode “Universal ASR” model \cite{yu2020universal}.

\vspace{-0.05in}

\bibliographystyle{IEEEbib}
\bibliography{refs}

\end{document}